\documentclass[12pt,letterpaper]{article}
\pdfoutput=1

\usepackage{amsmath,amssymb,calc, amsthm,bbm, epsfig,psfrag, mathtools}
\usepackage{graphicx, enumerate}
\usepackage[numbers,sort&compress]{natbib}
\usepackage[dvipsnames]{xcolor}

\usepackage[utf8]{inputenc} 

\newcommand{\Comment}[1]{{}}
\definecolor{darkblue}{rgb}{0.15,0.35,0.55}
\definecolor{reddish}{rgb}{0.65, 0.2, 0.2}
\usepackage[linktocpage=true]{hyperref}
\hypersetup{
colorlinks=true,
citecolor=darkblue,
linkcolor=reddish,
urlcolor=darkblue,
pdfauthor={},
pdftitle={},
pdfsubject={}
}

\makeatletter
\renewcommand\section{\@startsection {section}{1}{\z@}%
                                   {-3.5ex \@plus -1ex \@minus -.2ex}
                                   {2.3ex \@plus.2ex}%
                                   {\normalfont\large\bfseries}}
\renewcommand\subsection{\@startsection{subsection}{2}{\z@}%
                                     {-3.25ex\@plus -1ex \@minus -.2ex}%
                                     {1.5ex \@plus .2ex}%
                                     {\normalfont\bfseries}}
\makeatother

\theoremstyle{plain}

\theoremstyle{definition}

\let\non\nonumber

\def\bea#1\eea{\begin{align}#1\end{align}}

\def\bes #1\ees{\begin{split}#1\end{split}}

\newcommand{\be}{\begin{equation}}
\newcommand{\ee}{\end{equation}}

\newcommand{\bma}{\begin{pmatrix}}
\newcommand{\ema}{\end{pmatrix}}

\newcommand{\Z}{{\mathbb Z}}
\newcommand{\R}{{\mathbb R}}

\newcommand{\PP}{{\mathbb P}}

\let\l=\lambda

\let\s=\sigma

\def\e{\epsilon}

\def\l{\lambda}
\def\m{\mu}
\def\n{\nu}

\def\s{\sigma}

\def\M{{\mathcal M}}

\newcommand{\jbar}{{\bar \jmath}}

\newcommand{\p}{\partial}

\newcommand{\C}[1]{$(\ref{#1})$}

\def\IZ{\relax\ifmmode\mathchoice
{\hbox{\cmss Z\kern-.4em Z}}{\hbox{\cmss Z\kern-.4em Z}}
{\lower.9pt\hbox{\cmsss Z\kern-.4em Z}} {\lower1.2pt\hbox{\cmsss
Z\kern-.4em Z}}\else{\cmss Z\kern-.4em Z}\fi}
\def\IR{\relax{\rm I\kern-.18em R}}

\def\one{{\hbox{ 1\kern-.8mm l}}}

\def\tr{{\rm tr\,}}
\def\Tr{{\rm Tr\,}}

\newlength{\bredde}
\def\slash#1{\settowidth{\bredde}{$#1$}\ifmmode\,\raisebox{.15ex}{/}
\hspace*{-\bredde} #1\else$\,\raisebox{.15ex}{/}\hspace*{-\bredde}
#1$\fi}

\newsavebox{\zzzbar}
\sbox{\zzzbar}
  {\setlength{\unitlength}{0.9em}
  \begin{picture}(0.6,0.7)
  \thinlines
  \put(0,0){\line(1,0){0.6}}
  \put(0,0.75){\line(1,0){0.575}}
  \multiput(0,0)(0.0125,0.025){30}{\rule{0.3pt}{0.3pt}}
  \multiput(0.2,0)(0.0125,0.025){30}{\rule{0.3pt}{0.3pt}}
  \put(0,0.75){\line(0,-1){0.15}}
  \put(0.015,0.75){\line(0,-1){0.1}}
  \put(0.03,0.75){\line(0,-1){0.075}}
  \put(0.045,0.75){\line(0,-1){0.05}}
  \put(0.05,0.75){\line(0,-1){0.025}}
  \put(0.6,0){\line(0,1){0.15}}
  \put(0.585,0){\line(0,1){0.1}}
  \put(0.57,0){\line(0,1){0.075}}
  \put(0.555,0){\line(0,1){0.05}}
  \put(0.55,0){\line(0,1){0.025}}
  \end{picture}}

\newfont{\goth}{ygoth.tfm scaled 1200}                   

 \numberwithin{equation}{section}

\def\1{{(1)}}
\def\2{{(2)}}
\def\3{{(3)}}

\newcommand{\ul}{\underline}

\begin{document}
\begin{titlepage}

\begin{center}

September 12, 2017
\hfill         \phantom{xxx}  EFI-17-5

\vskip 2 cm {\Large \bf Supersymmetry Breaking by Fluxes} 
\vskip 1.25 cm {\bf Savdeep Sethi}\non\\
\vskip 0.2 cm
 {\it Enrico Fermi Institute \& Kadanoff Center for Theoretical Physics \\ University of Chicago, Chicago, IL 60637, USA}

\vskip 0.2 cm

\end{center}
\vskip 1.5 cm

\begin{abstract}
\baselineskip=18pt

Type II string theory and M-theory admit flux configurations that break supersymmetry below the Kaluza-Klein scale. These backgrounds play a central role in most models of the string landscape. I argue that the behavior of such backgrounds at weak coupling is generically a rolling solution, not a static space-time. Quantum corrections to the space-time potential are computed around this classical time-dependent background. This is particularly important for non-perturbative corrections. This change in perspective offers an explanation for why there appear to be many effective field theory models that seemingly evade the known no-go theorems forbidding de Sitter space-times. This has interesting implications for type IIB string landscape models.  

\end{abstract}

\end{titlepage}


\section{Introduction} \label{intro}

\begin{figure}
    \centering
    \begin{minipage}{0.45\textwidth}
        \centering
        \includegraphics[width=0.7\textwidth]{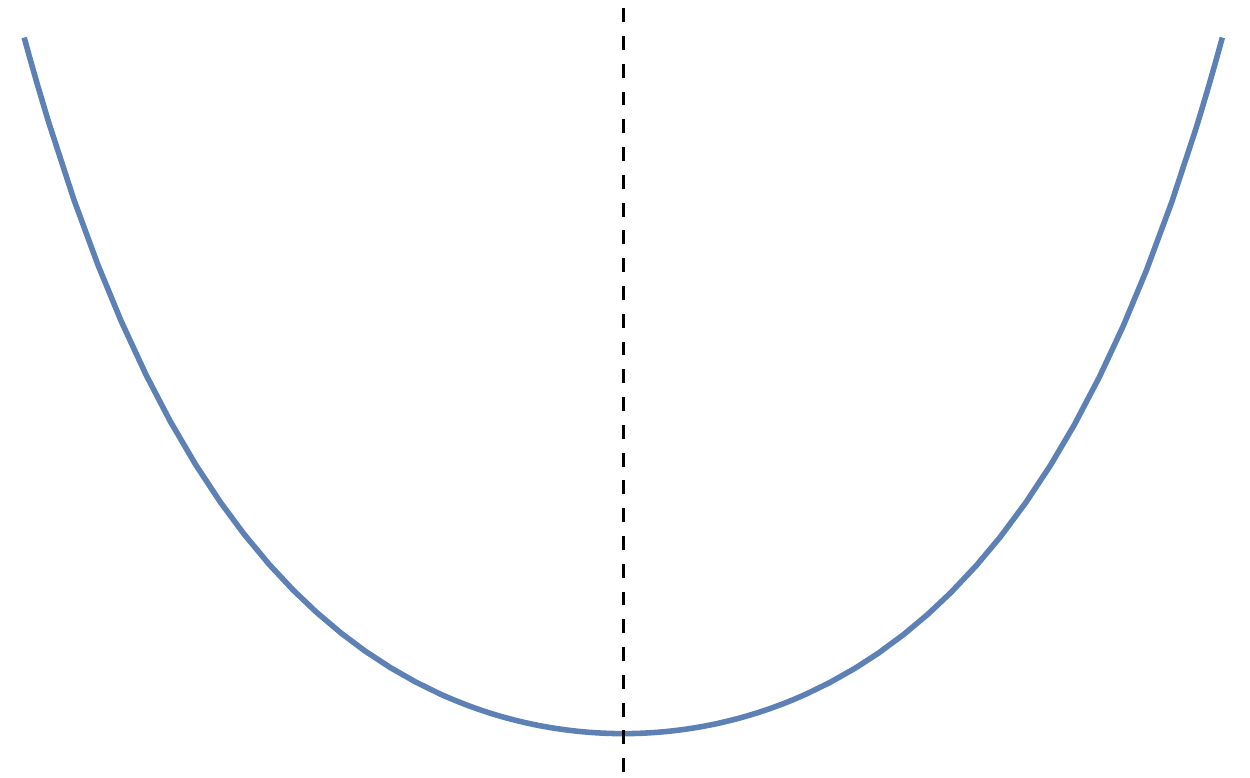} 
        \caption{A good starting point.} \label{figure1}
    \end{minipage}\hfill
    \begin{minipage}{0.45\textwidth}
        \centering
        \includegraphics[width=0.7\textwidth]{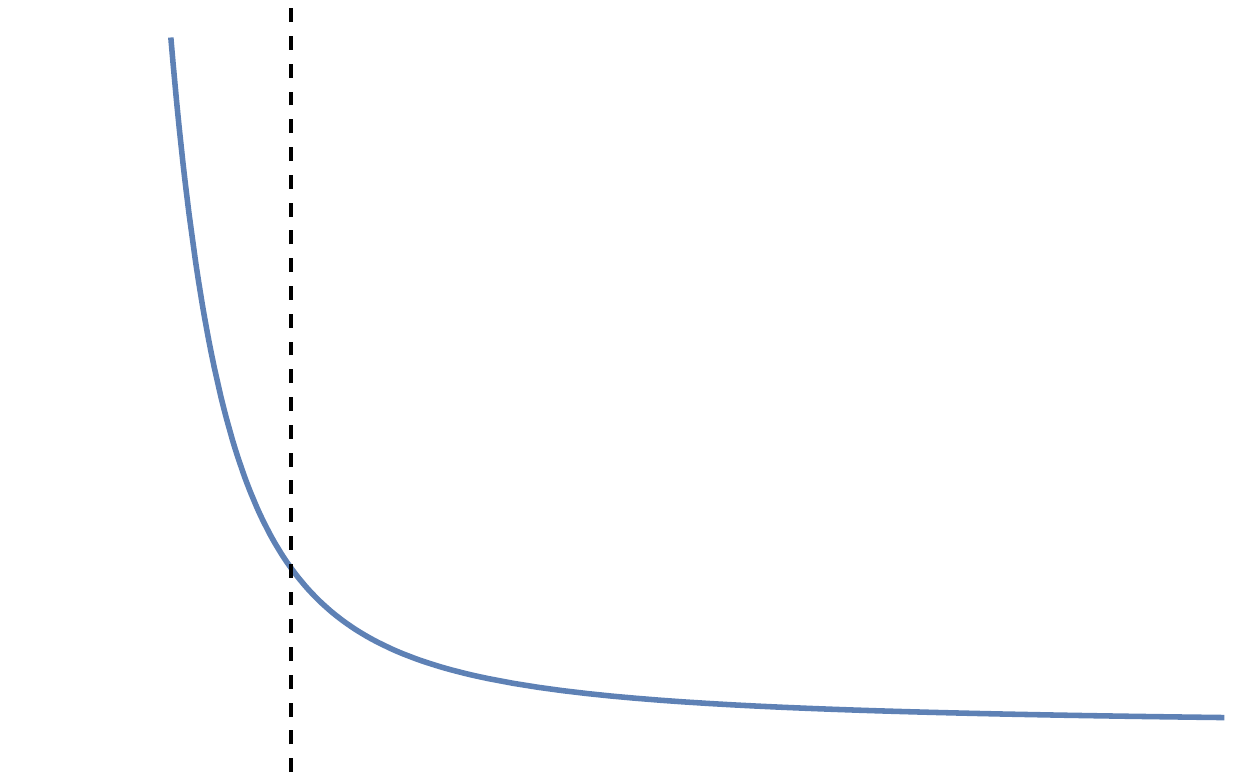} 
        \caption{A not so good starting point.} \label{figure2}
    \end{minipage}
\end{figure}

From the perspective of effective field theory, the construction of a de Sitter space-time would appear to be a simple task. One simply has to engineer a potential of the form displayed in figure~\ref{figure1}\ with positive energy at the minimum in a controlled model. This task, however, has proven to be extremely challenging in string theory.  The obstructions to building a de Sitter space-time in string theory are captured in a successive series of no-go results, which are worth overviewing: 
\begin{enumerate}[(i)]
\item
The fundamental supergravity theories that arise in D=10 and D=11 dimensions obey the strong energy condition (SEC). This condition is inherited by any smooth compactification of these theories~\cite{Gibbons:1984kp, Maldacena:2000mw}.\footnote{For constraints on higher-dimensional theories obeying the null energy condition with any form of accelerated expanion, see~\cite{Steinhardt:2008nk}.}  On the other hand, an accelerating universe requires a violation of the strong energy condition.\footnote{This does not mean that time-dependent backgrounds with a transient phase of acceleration are impossible in supergravity. If one relaxes the condition that the internal space is static then solutions like those described in~\cite{Townsend:2003fx, Emparan:2003gg}\ are possible.}

\item
String theory is not supergravity. There are exotic objects like orientifold planes that violate the SEC. These violations come from higher derivative interactions supported on the planes. At least for heterotic/type I theories, the leading SEC violating interactions still do not permit accelerating universes~\cite{Green:2011cn, Gautason:2012tb}. These interactions are, however, sufficient to allow compactifications to Minkowski space-time with non-vanishing internal fluxes. 
This result has been extended to include the effect of gaugino condensation, with a similar failure to find any accelerating solutions~\cite{Quigley:2015jia}.

While this analysis applies to the interactions supported on the O9-planes of type I string theory, it is highly likely that a similar result can be found for the lower-dimensional orientifold planes that appear in other constructions, largely because space-time supersymmetry determines much of the structure of the relevant couplings on the planes. Indeed in the context of type IIB compactifications, similar results have been seen in~\cite{Andriot:2016xvq, Dasgupta:2014pma, Junghans:2016uvg, Moritz:2017xto}.

\item
String theory is not just supergravity augmented with a collection of higher derivative interactions. There are quantum corrections which are non-perturbative both in the string length and in the string coupling.  At least in the heterotic string, one can rule out macroscopic de Sitter space-times of dimension four or higher at the level of classical string theory~\cite{Kutasov:2015eba}. This analysis captures strong curvature effects on the world-sheet beyond the reach of supergravity, including world-sheet instantons. Supersymmetry plays no role at all. 
\end{enumerate}

On the other hand, there are many proposals for four-dimensional de Sitter  and anti-de Sitter solutions in type IIB string theory based on an N=1 D=4 effective supergravity theory characterized by a K\"ahler potential $K$ and a superpotential $W$. In terms of these quantities, the physical potential is given by
\be\label{V}
V = e^K \left( K^{i\jbar} D_i W D_{\jbar}  \bar{W} - 3 |W|^2\right), \qquad D_i = \p_i + \p_i K.
\ee For the proposed dS and AdS constructions, $K$ and $W$ are usually of the following schematic type:
\be\label{basicdata}
K = -3 \log(\rho +\bar\rho), \qquad W=W_0 + A e^{-a \rho}. 
\ee
Here $\rho$ is a K\"ahler modulus, $W_0$ is a constant generated by fluxes and $A e^{- a \rho}$ depicts some generic instanton or other non-perturbative corrections to $W$. The potential for this effective field theory model stabilizes the K\"ahler modulus.  In this discussion, I will largely ignore other moduli like complex structure moduli whose stabilization is more straightforward~\cite{Dasgupta:1999ss, Gukov:1999ya}. 

The proposed models fall into three broad categories: the original KKLT construction~\cite{Kachru:2003aw}, which requires adding an anti-D3-brane to `uplift' a supersymmetric stabilized anti-de-Sitter background to a de Sitter background.\footnote{The bullish case for the KKLT scenario is nicely described in~\cite{Polchinski:2015bea}.} 
 The large volume scenario which includes a perturbative correction to the K\"ahler potential of~\C{basicdata}, and some uplifting mechanism~\cite{Balasubramanian:2005zx}. In this case, both non-supersymmetric AdS and dS solutions are claimed to exist at parametrically large volumes compared to the string scale. Finally, the K\"ahler uplift scenario which involves no extra ingredients beyond perturbative corrections to $K$ and non-perturbative corrections to $W$ to produce dS solutions~\cite{Westphal:2006tn, Rummel:2011cd}.  The K\"ahler uplift scenario is, perhaps, the most surprising  model because it really involves no peculiar ingredients, but just arguments about the generic form of quantum corrections to the effective field theory action.

There is no sharp contradiction between the no-go results (i)--(iii) and these proposed constructions. None of the no-go results is sufficient to rule out de Sitter solutions from standard string compactifications. In each case, there is some quantum effect that is missed, which could well be the key ingredient needed for a de Sitter solution. What they do indicate is that any construction of a de Sitter space-time must involve an unconventional, or at least hard to compute ingredient. For example, orientifolds do not appear to be sufficiently exotic to evade the de Sitter no-go results. They are sufficiently exotic to permit  flux compactifications to Minkowski space-time, evading the supergravity no-go result (i)~\cite{Dasgupta:1999ss}. Certainly anti-branes are not exotic in any sense from the perspective of their stress-energy properties. Even in conjunction with orientifold planes, anti-branes are not expected to give de Sitter solutions.\footnote{ There has been a vibrant discussion about whether anti-branes are sensible as uplift ingredients, and whether the singularities they produce in supergravity can be interpreted sensibly; see~\cite{Cohen-Maldonado:2015ssa, Bena:2016fqp, Danielsson:2016cit, deAlwis:2016cty}\ for recent discussions. This topic is fascinating, but a digression from our present discussion which involves more elementary issues.}

One way of reconciling the ease with which one can construct effective field theory models of de Sitter with the no-go theorems is that something significant is missed in the no-go results. This is not a very satisfying answer because the effective field theory constructions claim to describe string backgrounds in regimes of very large volumes and weak string coupling, where we might expect the quantum corrections missed by the no-go results to be insignificant.\footnote{ Similar comments apply to the construction of stabilized anti-de Sitter space-times with the AdS length scale significantly larger than the Kaluza-Klein scale. There is a supergravity no-go result in this case as well~\cite{Gautason:2015tig}, extending no-go result (i). While scale separated AdS backgrounds emerge easily from effective field theory scenarios, no explicit 
constructions are known at this time.}  In this work I will propose a different explanation. If there is a universal issue with the effective field theory constructions then it should afflict all such models, not just one class of models or another, and we will see that this is indeed the case.   The key issue 
is the interplay between supersymmetry breaking, time-dependence and instanton corrections. 


\subsection{Setting expectations}


The effective field theory models of~\C{basicdata}\ are based on type IIB flux compactifications. Supersymmetry preserving type IIB flux compactifications were first constructed by DRS~\cite{Dasgupta:1999ss}. I will clarify what I mean by ``constructed'' momentarily. The type IIB supergravity $3$-form flux $G_3$ is required to be imaginary self-dual (ISD). For the case of Calabi-Yau 3-fold orientifold compactifications, this means it has Hodge type $(2,1)$ with no $(0,3)$ component.  An extension of this construction to flux backgrounds that break supersymmetry was subsequently proposed by GKP~\cite{Giddings:2001yu}. In this case, the flux includes components of Hodge type $(0,3)$. I want to  contrast these two cases because this difference is the heart of the issue.

There are three related Minkowski space-time backgrounds that really should be considered together. The basic input data in each case is a choice of Calabi-Yau $4$-fold $\M$ and a choice of flux. The three backgrounds are:
\begin{enumerate}[(A)]
\item
 D=2; Type IIA on $\M$,
 \item
 D=3; M-theory on $\M$, 
 \item
 D=4; F-theory on $\M$. 
\end{enumerate}
To have a Lorentz invariant F-theory or type IIB background requires that $\M$ should be elliptically fibered with section~\cite{Vafa:1996xn}, and there are further restrictions on the choice of flux~\cite{Dasgupta:1999ss}. For cases (A) and (B), there is no restriction on the geometry $\M$. These three cases are classically related by compactification: 
\be\label{relatedback}
\text{ D=4; F-theory on $\M$}  \, \xrightarrow{\times S^1} \, \text{ D=3; M-theory on $\M$} \,\xrightarrow{\times S^1} \, \text{ D=2; type IIA on $\M$}. 
\ee
Although these three backgrounds are related by a circle compactification, in each case the relation really involves strong-weak coupling. This is particularly important for compact flux solutions because there are no classical solutions at the level of supergravity as a consequence of no-go result (i). The only current method we have for studying compact flux solutions is using the space-time equations of motion derived from the space-time effective action for either type IIA string theory, M-theory or type IIB string theory. As soon as one needs higher derivative interactions beyond supergravity, the relation between these three space-time effective actions is no longer classical, but involves quantum corrections.

The most poorly understood of the three cases is the F-theory background because the couplings supported on O7-planes and D7-branes, or more generally $(p,q)$ 7-branes, are not completely known even at low orders in the derivative expansion. The M-theory and type IIA backgrounds are better understood, but even there our knowledge of the space-time effective actions is incomplete even for the leading higher derivative interactions.   

It is a common belief that the existence of flux vacua is well established. Nothing could be further from the truth! To see why, it is first worth recalling why a $(2,2)$ sigma model with a Calabi-Yau target space defines a two-dimensional $(2,2)$ conformal field theory perturbatively in $\alpha'$~\cite{Nemeschansky:1986yx}. The data defining the sigma model is a choice of K\"ahler potential for the target manifold. In $(2,2)$ superspace with chiral superfields $\Phi$, the two-dimensional action takes the form:
\be
S = \int d^2x d^2\theta d^2\bar\theta \, K(\Phi, \bar\Phi). 
\ee 
Vanishing of the $1$-loop beta-function for such a sigma model requires a Ricci-flat target metric. The structure of $(2,2)$ superspace then guarantees that any higher sigma model loop contribution to the beta-function corresponds to a shift of the K\"ahler potential by a globally defined function. Such contributions can therefore be compensated by small modifications of the metric away from Ricci flat, which still preserve the cohomology class of the K\"ahler form.   

This classic world-sheet argument has no analogue for flux compactifications. This is because of the absence of a useful world-sheet description for backgrounds with RR flux. However, the world-sheet argument above can be easily reformulated as a space-time argument. Take the case of a Calabi-Yau $3$-fold and the heterotic string for specificity. A Ricci-flat metric solves the equations of motion at lowest order in $\alpha'$. Identifying the gauge bundle with the tangent bundle, appropriate for a world-sheet $(2,2)$ model, solves the equations of motion at order $\alpha'$ and satisfies the heterotic Bianchi identity.  

The resulting D=4 space-time effective field theory has N=1 supersymmetry and is again characterized by a space-time $(K,W)$. There is no tree-level $W$ because this background preserves supersymmetry to $O(\alpha')$. Generating a space-time potential requires either a $W$ or an F-I term. However, symmetry in the form of holomorphy and scaling arguments now guarantee that no higher loop in sigma model perturbation theory can generate either coupling. The background can still be destabilized by world-sheet instantons, and by both perturbative and non-perturbative corrections in the string coupling. Regardless, there exists a static starting configuration around which one can systematically compute quantum corrections. This kind of space-time reasoning has recently been applied in other settings~\cite{Becker:2014rea, Becker:2015bra}.  

The key point relevant for flux vacua is that a $W_0$, generated by a supersymmetry breaking flux, presents an obstruction to the existence of a solution to the  perturbative space-time equations of motion.  When supersymmetry is preserved and $W_0=0$, there is a fighting chance that solutions to the equations of motion at low orders in the momentum expansion can be promoted to full solutions of the space-time equations of motion; there are no obvious obstructions. It is in this sense that supersymmetric type IIB flux solutions exist. 

When $W_0\neq 0$, there is no good reason to expect that a perturbative solution exists. In fact, the opposite is true. Generically, one should never expect a static solution since any modification of the no-scale K\"ahler potential of~\C{basicdata}\ is likely to generate a physical potential of perturbative strength in $\ell_p$. This is why the DRS and GKP backgrounds are on such different footing. 

There are really only three points we need to understand. The first is the role that higher derivative interactions play in evading no-go result (i) and permitting compact flux solutions. The second point is whether corrections to the $K$ of~\C{basicdata}\ are likely from higher derivative interactions, though intuitively this should seem completely reasonable. The final point is the computation of quantum corrections for a background with broken supersymmetry. 

Rather than discuss type IIB or F-theory flux vacua, let us first continue this analysis in class (B) consisting of M-theory compactifications to three dimensions~\cite{Becker:1996gj}. All of the issues we want to address are identical. The advantage of focusing on the D=3 M-theory backgrounds is that the complications involving the branes and orientifold planes of F-theory are replaced by questions about the higher derivative interactions in the M-theory space-time effective action.  The setting is cleaner and the physics more readily apparent. We will come back to the relation between the different classes of backgrounds a little later.  




\subsection{Point 1: Higher derivative terms} \label{higherderint}

The starting point for any construction of an M-theory solution is some background that solves the D=11 equations of motion derived from the space-time effective action. This action includes interactions at all orders in the derivative expansion. 
We assume a D=11 space-time manifold that is a product of D=3 Minkowski space with $\M$. The metric is generally warped, but respects the isometries of Minkowski space. In addition to the choice of manifold $\M$, we require a choice of $4$-form flux $G_4$ to complete our input data. The M-theory space-time effective action is given by $11$-dimensional supergravity at the two derivative level:
\be\label{11dsugra}
S_2 = {1\over 2\kappa_{11}^2} \int d^{11}x \sqrt{-G} \left( R - {1\over 2} |G_4|^2\right)  - {1\over 12\kappa_{11}^2} \int C_3\wedge G_4 \wedge G_4 +\ldots. 
\ee 
The omitted terms involve fermions. Let us define the $11$-dimensional Planck scale, $\ell_p$, so that $ {2 \kappa_{11}^2}=(2\pi)^8 \ell_p^9$ and the tension of an M2-brane is given by:
\be T_{M2} = {1\over (2\pi)^2 \ell_p^3}. \ee 
In the conventions of~\C{11dsugra}, the components of $C_3$ are dimensionless while the components of $G_4$ have mass dimension $1$. This is the natural normalization for the derivative expansion of the space-time effective action.

There are no compact flux solutions if we truncate the M-theory space-time effective action at two derivatives. As first realized in~\cite{Becker:1996gj}, which initiated the study of M-theory and type IIB flux vacua, we must include interactions with at least $8$ derivatives in the space-time effective action.   The $8$ derivative space-time couplings are all related by space-time supersymmetry, and $\ell_p^6$ suppressed compared with the supergravity couplings of~\C{11dsugra}. We can group the bosonic couplings in the following way:
\be\label{higherder}
S_8 =  {1\over (2\pi)^6  3^2 2^{13} \ell_p^3}\int \sqrt{-G} \left(t_8 t_8 - {1\over 24} \e_{11} \e_{11} \right) R^4 - T_{M2} \int C_3 \wedge X_8 + O( \left[G_4\right]^2). 
\ee
The omitted $O( \left[G_4\right]^2)$ terms refer to $8$ derivative couplings with at least $2$ $G_4$-flux operators. The couplings appearing in~\C{higherder}\ are given explicitly by
\be 
X_ 8 = {1\over (2\pi)^4}{1\over 192} \left( \Tr R^4 - {1\over 4} (\Tr R^2)^2  \right), \qquad \int_{\M} X_8 = - {\chi(\M) \over 24},
\ee
and by
\bea
\e_{11} \e_{11} R^4 &= \e^{B_1B_2B_3M_1\ldots M_8}\e_{B_1B_2B_3N_1\ldots N_8} R^{N_1N_2}_{M_1M_2} \cdots R^{N_7N_8}_{M_7M_8}, \cr
t_8t_8 R^4 & = t_8^{M_1\ldots M_8} t_{8 \, N_1\ldots N_8} R^{N_1N_2}_{M_1M_2} \cdots R^{N_7N_8}_{M_7M_8}, 
\eea
using the conventions of~\cite{Tseytlin:2000sf, Haack:2001jz, Grimm:2014xva, Grimm:2014efa, Grimm:2015mua}. Here $\e_{11}$ is the $11$-dimensional antisymmetric tensor, while $t_8$ satisfies 
\be
t^{M_1\ldots M_8} A_{M_1M_2}\cdots A_{M_7 M_8} = 24 \tr A^4 - 6(\tr A^2)^2,
\ee 
for antisymmetric tensors $A$. The combination $\e_{11}\e_{11} R^4$ is proportional to the $8$-dimensional Euler density satisfying,
\be\label{eeR4}
{\chi(\M)}={1\over (2\pi)^4 \cdot 3^2 \cdot 2^{12} } \int_{\M} \sqrt{g} \left( \e_{11}\e_{11} R^4 \right) , 
\ee
where $g$ refers to the metric of $\M$~\cite{Haack:2001jz}. 

The tadpole condition for M2-brane charge reads, 
\be\label{tadpole}
{1\over 2} \int_\M {{\widehat G_4}\over 2\pi} \wedge {{\widehat G_4}\over 2\pi}= {\chi(\M) \over 24} ,  \qquad {\widehat G_4} = T_{M2} \, G_4,
\ee
where we omit the possibility of adding explicit brane sources described in~\cite{SVW}. With this choice of normalization, flux quantization reads
\be
 \left[{{\widehat G_4}\over 2\pi} \right] - {p_1(\M) \over 4}  \in H^4(\M, \Z), 
\ee
where $p_1$ is the first Pontryagin class~\cite{Witten:1996md}.
The $O(\left[G_4\right]^2)$ terms in~\C{higherder}\ with $8$  derivatives are partially known~\cite{Hyakutake:2007sm, Liu:2013dna, Grimm:2014efa}, with even less known about terms with more flux couplings.

 It will be useful to estimate the size of terms. Let $\ell_\M$ be the characteristic size of the compact space $\M$. By assumption, $\ell_\M >> \ell_p$ otherwise the M-theory space-time effective action is not a useful starting point. Flux quantization implies that the internal flux is of size,
 \be G_4 \sim \ell_p^3/\ell_\M^4, \ee 
 so terms in the effective action involving larger numbers of $G_4$ fluxes are smaller by powers of $\ell_p^3$. This does not mean they are unimportant! For example, the $O(\left[G_4\right]^2)$ terms in~\C{higherder}\ are critical if one is to write a complete D=11 space-time supersymmetric action. 
 
Let $g^{(2)}$ denote the Ricci-flat metric for $\M$ with coordinates $y$, and let $\eta$ denote the D=3 Minkowski metric. Take a D=11 metric ansatz:
 \be\label{metricansatz}
 ds^2_{D=11} = e^{-{\cal W}(y)} \eta + e^{{1\over 2} {\cal W}(y)} \left( g^{(2)}_{ij} + g^{(8)}_{ij} + \ldots \right) dy^i dy^j.  
 \ee
In terms of notation, the indices $M,N,\ldots$ refer to the full $11$ dimensions, the indices $i,j,\ldots$ refer to internal directions, while the indices $\m,\n,\ldots$ refer to space-time directions. Expressing the metric in this way using a warp factor ${\cal W}(y)$ is a little bit of a historical carry over from the original analysis of~\cite{Becker:1996gj}.  The warp factor ${\cal W}$ itself has an expansion in powers of $\ell_p$. The correction $g^{(8)}$ reflects modifications of the Calabi-Yau metric from terms in the effective action~\C{higherder}\ with $8$ derivatives that cannot be absorbed into the warp factor ${\cal W}$. To this order in the momentum expansion, the metric is often stated to be conformally Calabi-Yau, but a recent analysis argues that $g^{(8)}$ is non-vanishing and deforms the metric away from the Ricci-flat choice~\cite{Grimm:2014xva}. There is no known reason for the exact metric solving the space-time equations of motion to be exactly conformally Ricci-flat so this seems quite reasonable. 

The last ingredient is a space-time filling $C_3$ potential given by the warp factor and the antisymmetric tensor of D=3 Minkowski space,
\be\label{sptimepot}
C_{\m\n\l}=  \e_{\m\n\l} e^{-{3\over 2} {\cal W}}.
\ee 
 The flux equations of motion, ignoring any eight derivative corrections, imply that the warp factor ${\cal W}$ is a constant up to terms of $O(\ell_p^6/\ell_\M^6)$.  The  field strength associated to the space-time filling $C_3$ potential is of order $O(\ell_p^6/\ell_\M^7)$, which therefore generates a stress-energy contribution of $O(\ell_p^{12}/\ell_\M^{14})$ that sources the Ricci tensor:
 \be
 R_{MN} = {1\over 12} \left( G_{MPQR} G_{N}^{PQR} - 2 g_{MN} |G|^2 \right). 
 \ee 

The first approximate solution to the equations of motion is a purely Ricci-flat metric with no flux. This is a not an actual solution to the space-time equations of motion unless $\chi(\M)=0$ precisely because of the tadpole constraint~\C{tadpole}. The next approximation requires the introduction of $G_4$-flux and, necessarily, the consideration of $8$ derivative couplings in the space-time effective action. Essentially all prior discussions have focused exclusively on the couplings appearing in~\C{higherder}\ basically because much less is known about the remaining couplings. Just at the level of solving the supergravity equations of motion, we require a flux that satisfies
\be
G_4= \ast_8 \, G_4.  
\ee
Here the Hodge dual is taken with respect to the internal metric on $\M$.  This is only strictly true if we neglect corrections to the flux equation of motion from the $8$ derivative terms. Each of those corrections is small and presents no intrinsic obstacle to finding a solution. If we insist on preserving $4$ real supercharges then supergravity requires the following constraints,
\be
J\wedge G_4=0, \qquad G_4 \in H^{(2,2)}(\M), 
\ee
where $J$ is the K\"ahler form for $\M$. This is the situation that has been most heavily studied because it can lift to a supersymmetric F-theory background, assuming appropriate conditions on the flux are satisfied~\cite{Dasgupta:1999ss}. In this situation the constant $W_0$ of the effective D=3 space-time theory~\C{basicdata}\ vanishes. There are more general flux configurations that preserve $2$ or more real supercharges explored in~\cite{Prins:2013wza, Melnikov:2017wcf}\ for which the flux is neither purely of $(2,2)$ Hodge type,  nor primitive. We will not need those more general cases in this discussion. 

The case of most interest for us is when the metric is chosen so that the flux breaks supersymmetry because it has components of Hodge type $(4,0)$ and $(0,4)$. Such supersymmetry breaking backgrounds have been discussed in~\cite{Becker:2002nn, Berg:2002es}. In this case $W_0$ of the GVW superpotential is activated. The relevant term in the GVW superpotential is expressed in terms of the holomorphic $4$-form $\Omega$ of $\M$~\cite{Gukov:1999ya}:
\be\label{GVW}
W = \int G_4 \wedge \Omega + \ldots.
\ee
In deriving the superpotential~\C{GVW}\ from Kaluza-Klein reduction on the background~\C{metricansatz}, the effect of the $8$ derivative pure metric coupling of~\C{higherder}\ plays a critical role~\cite{Haack:2001jz}. When evaluating that coupling to leading order in an $\ell_p$ expansion, we can ignore all the corrections to the basic metric $g^{(2)}$ and ignore all warping and fluxes, because that coupling is already suppressed by $\ell_p^6$.  Evaluated directly on the internal metric, that metric coupling produces a physical potential for the overall volume from~\C{eeR4}\ together with the observation that 
\be
\int \sqrt{g} \left( t_8 t_8+ {1\over 24} \e_{11}\e_{11} \right) R^4 = 0, 
\ee
when integrated over $\M$~\cite{Gross:1986iv}. This potential  cancels a similar contribution from the $|G_4|^2$ flux kinetic term in~\C{11dsugra}\ if the tadpole cancelation condition~\C{tadpole}\ is satisfied. 

For this reason, there is no sense in which the $8$ derivative metric coupling of~\C{higherder}\ is negligible, even at large volume. It is as important as the supergravity flux couplings. However to write a complete D=3 space-time supersymmetric effective theory, one is then forced to consider all the remaining $8$ derivative terms which are related to the $R^4$ terms by supersymmetry. For example, the $O(\left[G_4\right]^2)$ terms in~\C{higherder}\ are suppressed by $\ell_p^{12}$,  which is the same order as the stress-energy produced by the space-time $C_3$ potential~\C{sptimepot}. These terms again cannot be neglected, though less is known about those couplings. At the $8$ derivative level, the most suppressed possible bosonic coupling in the supersymmetric completion of $R^4$ is $O(\left[G_4\right]^8)$, which down by a factor of $\ell_p^{30}$. 

This logic extends 
to higher orders in the derivative expansion. 
The next set of higher derivative interactions,  beyond $8$ derivatives, are expected to involve $14$ derivatives. Schematically, these are $O(R^7)$ couplings suppressed by $\ell_p^{12}$ relative to the supergravity interactions, but this is the same order as the $8$ derivative $O(\left[G_4\right]^2)$ terms. These terms can also generate corrections to the space-time $K$ and so cannot be ignored when $W \neq 0$. 

This is point one: to construct a compact flux solution in M-theory, all orders in the momentum expansion are relevant.  There is no simple truncation of the M-theory space-time effective action to a finite subset of the higher derivative interactions, whose study alone can ensure the existence of a solution. This is particularly critical for backgrounds that spontaneously break supersymmetry; the non-vanishing flux superpotential is an obstruction to the existence of a classical static solution, which cannot be compensated by any perturbative modification of the background.  

\subsection{Higher derivative couplings in type IIA and F-theory}

We can now try to make similar statements in the related type IIA and F-theory backgrounds of~\C{relatedback}. These backgrounds are related by strong-weak coupling duality. What allows us to say something about the backgrounds and the relevant space-time effective actions is supersymmetry. Maximal supersymmetry fixes not only the supergravity interactions, but also at least the $8$ derivative interactions. First let us consider the relation between M-theory on $\M\times S^1$ and type IIA on $\M$. Take $S^1$ to have size $R_{11}$. The D=10 type IIA analogue of the D=11 $R^4$ couplings in string frame take the form,\footnote{These type II couplings are found in many papers. A nice summary with relevant references can be found in~\cite{Hyakutake:2006aq}\ based in part on~\cite{Tseytlin:2000sf}.} 
\bea\label{typeIIAR4tree}
S_8^{\rm IIA} =  {1\over 3 \cdot 2^{11} \cdot (2\pi)^7 \alpha'}&  \int d^{10}x \sqrt{-G}  \left\{   \zeta(3) e^{-2\phi} \left( t_8 t_8 + {1\over 8} \e_{10} \e_{10} \right) R^4 + \right.  \\  & \left.    {\pi^2\over 3}\left[ \left( t_8 t_8 - {1\over 8} \e_{10} \e_{10} \right) R^4  - {1\over 2} \e_{10} t_8 B_2 R^4\right] \right\} +\ldots. \label{linetwo}
\eea
Here $B_2$ is the NS $B$-field with the  $B_2 R^4$ coupling descending from $C\wedge X_8$ of~\C{higherder}. The string coupling is related to $11$-dimensional quantities as usual by $e^{\phi} = \left( R_{11} \over \ell_p \right)^{3/2}$.  The first line~\C{typeIIAR4tree}\ is not obtained by straight dimensional reduction of the M-theory couplings~\C{higherder}; rather those terms are generated at tree-level in type IIA string theory. From a space-time perspective, the string tree-level term is generated by a loop of D=11 gravitons on $S^1$~\cite{Green:1997as}. The terms that follow from D=11 by dimensional reduction are generated at one string loop and appear on the second line~\C{linetwo}. The moduli-dependence of the $R^4$ couplings is determined by supersymmetry so there are no further loop or non-perturbative corrections~\cite{Green:1998by}. 

The same scaling arguments presented in section~\ref{higherderint}\ apply directly to this case. The supergravity couplings involving the $G_4$ internal flux mix with the $R^4$ terms, and consistency then requires the inclusion of all higher derivative couplings. There is again no way to truncate the space-time effective action to a finite subset of the higher derivative interactions. However in this string theory setting, there is a new parameter determined by the string coupling, or the size of the circle $R_{11}$. One could organize a perturbative construction of a space-time solution order by order in the string loop expansion. Each order in the loop expansion still involves an infinite tower of higher derivative interactions! For this to be a practical approach, however, requires a useful  world-sheet formalism for RR fluxes.  

Note that simply dimensionally reducing a D=3 solution of M-theory on $S^1$ does not generically give a solution to the type IIA equations of motion because of the new couplings in~\C{typeIIAR4tree}\ along with new string tree-level interactions found at higher orders in the momentum expansion. Fortunately at the $8$ derivative order, the integral of the combination $\left( t_8 t_8 + {1\over 8} \e_{10} \e_{10} \right) R^4$ over a Ricci-flat K\"ahler manifold vanishes. Otherwise, this coupling would already produce a non-vanishing potential for the dilaton and volume modulus even if tadpole cancelation is satisfied. 

Finally let us turn to F-theory on an elliptically fibered $\M \rightarrow B_6$ with base $B_6$. A quantum treatment of a generic F-theory background is difficult because of both high curvatures and the absence of a tunable string coupling. For the point we would like to explore, it is sufficient to restrict to models which admit an orientifold locus, where the string coupling is tunable in the absence of fluxes~\cite{Sen:1997bp}. In the presence of type IIB $G_3$ fluxes, the string coupling is frozen for orientifold models, along with generically all the remaining complex structure moduli~\cite{Dasgupta:1999ss}. On this locus, F-theory on $\M$ is equivalent to an orientifold compactification of type IIB on a Calabi-Yau $3$-fold $\widetilde\M$ determined by $B_6$. 

There are two sets of higher derivative interactions found in this background. There are bulk type IIB couplings as well as couplings supported on the O7-planes and D7-branes. The leading bulk $R^4$ terms are again fully determined by supersymmetry. We only need the terms perturbative in the type IIB string coupling, which take the form:
\bea\label{typeIIBR4tree}
S_8^{\rm IIB} =  {1\over 3 \cdot 2^{11} \cdot (2\pi)^7 \alpha'}&  \int d^{10}x \sqrt{-G}  \left\{   \zeta(3) e^{-2\phi} \left( t_8 t_8 + {1\over 8} \e_{10} \e_{10} \right) R^4 + \right.  \\  & \left.    {\pi^2\over 3} \left( t_8 t_8 + {1\over 8} \e_{10} \e_{10} \right) R^4  \right\} +\ldots. \label{linetwoB}
\eea
In this case, both the string tree-level  and one-loop terms in~\C{typeIIBR4tree}\ and \C{linetwoB}\ are visible by studying the contribution to the $4$ graviton scattering amplitude from a space-time loop of 11-dimensional gravitons on the elliptic fiber of $\M$, for the case where the fibration is trivial~\cite{Green:1997as}.

The tree-level $R^4$ terms of~\C{typeIIBR4tree}\ are the leading higher derivative interactions in a string loop expansion. However, the string coupling is not a good expansion parameter for type IIB flux vacua because it cannot be tuned to parametrically small values. The choice of flux determines a value for the string coupling~\cite{Dasgupta:1999ss}. For some examples, it might be small but a controlled perturbative expansion really requires a parametrically tunable coupling. The only tunable expansion parameter in this type IIB orientifold model is the volume of $\widetilde\M$. 

In addition to the bulk couplings, there are higher derivative interactions supported on the O7-planes and D7-branes.  From the disk-level action, one finds $R^2$ type couplings of the form 
\be\label{Rsquared}
{1\over \ell_s^4} \int e^{-\phi} \, \tr  \left( R\wedge \ast R\right), 
\ee
as well as the coupling proportional to, 
\be\label{D3anomaly}
{1\over \ell_s^4} \int C_4\wedge {1\over 8\pi^2}\tr  \left(  R\wedge R \right), 
\ee
which plays a key role in canceling the D3-brane charge produced by the background $G_3$-flux. These couplings are supplemented by a host of interactions, required by supersymmetry, involving $G_3$-flux such as $\left( \nabla G_3 \right)^2$ terms. Flux quantization implies that the internal components of the $G_3$-flux are of order
\be
G_3 \sim \ell_s^2/\ell_{\widetilde \M}^3, 
\ee
ignoring factors of $g_s$ that differentiate the NS and RR components of $G_3$. Therefore couplings like $\left( \nabla G_3 \right)^2$ are further suppressed by $\ell_s^4$ when compared with terms like~\C{Rsquared}\ or~\C{D3anomaly}. 

There are also higher derivative brane supported interactions that can modify the $R^4$ couplings of~\C{typeIIBR4tree}\ and~\C{linetwoB}. Indeed the complete effective action supported on the planes and branes is not completely known, even at low orders in the derivative expansion; see~\cite{GarciaEtxebarria:2012zm, Grimm:2013gma, Grimm:2013bha, Minasian:2015bxa}\  for progress on determining these couplings. Fortunately the detailed form of the effective action will not be very important for us.  The key point again is that the existence of a solution to the space-time equations of motion that breaks supersymmetry cannot be determined by simply studying supergravity together with selected higher derivative terms that solve the tadpole constraint.   





\subsection{The K\"ahler potential}

Ideally to check if we have a static solution to the space-time equations of motion, we would directly compute the physical space-time potential generated by the choice of metric and flux. 
 This is a difficult task because the potential can be generated by higher derivative interactions involving fluxes, whose form is currently unknown. This is true in each of the three related backgrounds of~\C{relatedback}.  
Instead of directly computing the potential, we can examine corrections to $K$ and use~\C{V}\ together with the perturbative non-renormalization of $W$ to determine the space-time potential. The only assumption is that the low-energy theory can be written in a manifestly supersymmetric form. This is not a trivial assumption if there is no static solution. In a cosmological setting, the question of whether the low-energy effective theory has any superspace presentation will typically have a time-dependent answer. Operating under this assumption, the germane question then becomes not so much whether $K$ is corrected, since $K$ is not a protected quantity, but at what order in the $\ell_p$ expansion to expect the first correction.


Although this question is seemingly easier to address than directly computing the potential, it is still rather difficult because it involves finding both the functional form of $K$ and the correct complex coordinates for the K\"ahler moduli space. On the other hand, the physical potential does not care about the choice of parameterization of the scalar fields: either it vanishes or it does not vanish. 

The classical Calabi-Yau $3$-fold K\"ahler potential is given in terms of the volume ${\cal V}$, 
\be
K = -2 \log({\cal V}), 
\ee
but this expression is modified by both warping and fluxes, and by higher derivative interactions.  Even truncating the space-time effective action to $8$ derivatives, there are only incomplete results to date. Most analyses have either taken into account warping and ignored higher derivative interactions, or taken into account higher derivative interactions and ignored warping. I will summarize the current data on the K\"ahler potential. 

If one accounts for warping and fluxes at the level of two derivative type IIB supergravity then there are compelling arguments that the K\"ahler coordinates are corrected, but that the K\"ahler potential retains its no-scale property when expressed in terms of the corrected coordinates~\cite{Frey:2013bha, Martucci:2014ska, Martucci:2016pzt}. The main point about the new coordinates is that they depend on the choice of fluxes, and the choice of complex structure and brane moduli. The corrected K\"ahler potential is still given in terms of the volume, but in terms of the warped volume ${\widehat {\cal V}}$ rather than the unwarped volume
${\cal V}$:
\be
K = - 2 \log({\cal V}) \rightarrow -2 \log({\widehat {\cal V}}). 
\ee



On the other hand, there is considerably more data available if one ignores fluxes and warping, and purely studies the Calabi-Yau geometry. 
We can first ask about the structure of quantum corrections to the K\"ahler potential for the K\"ahler moduli space of a $(2,2)$ world-sheet theory with target space $\M$. This would be the world-sheet theory for the type IIA string compactified to D=2 on $\M$. As a consequence of the M-theory tadpole noted earlier, such a type IIA compactification has a tadpole for $B_2$, but the tadpole only appears at one string loop, while the K\"ahler potential already receives quantum corrections at tree-level in string perturbation theory. 


The exact K\"ahler potential at large radius for such a world-sheet theory has a beautiful conjectured form based on evidence from explicit localization computations~\cite{Honma:2013hma, Halverson:2013qca}.  If $J_k$ form a basis for $H^{(1,1)}(\M)$  and $t^k$ are complex coordinates on the K\"ahler moduli space then 
\bea \label{kahler4}
 & e^{-K(t)} =   {1\over 4!} \kappa_{ijkl}(t^i - \bar t^i) (t^j - \bar t^j) (t^k - \bar t^k) (t^l - \bar t^l)  + {i \over 4 \pi^3} \zeta(3) {\alpha'}^3  (t^k - \bar t^k) \times \cr & \int_\M J_k \wedge c_3 + O(e^{2\pi i t}), \\
& \kappa_{ijkl} = \int_\M J_i\wedge J_j\wedge J_k\wedge J_l,
\eea
%
where 
$c_3$ is the third Chern class of $\M$. The world-sheet instanton corrections are not relevant for our discussion, but the $\zeta(3) {\alpha'}^3$ term is highly relevant. This is the only correction perturbative in $\alpha'$ with coefficient determined by:
\be
\int_\M J_k\wedge c_3. 
\ee
It is precisely of the order we expect from corrections induced by the space-time higher derivative couplings~\C{typeIIAR4tree}. Before proceeding, let us check that it is non-vanishing in a simple example. Take the sextic CY 4-fold given by $\sum_i z_i^6 \subset \PP^5$. There is a unique K\"ahler class $J$ and a single $t$ coordinate. Adjunction gives us the Chern classes and the value of the perturbative correction to K, 
\be
c(\M) = {(1+J)^6 \over (1+6J)}, \qquad \int_\M J\wedge c_3 = -420. 
\ee
The point is that it does not vanish, and there is no reason to expect it to generically vanish. 

We can also check that this quantum correction breaks the no-scale condition for this theory:
\be
K^{i\jbar} K_i K_{\jbar} = 4. 
\ee
For simplicity, let us check this condition in a model with a single K\"ahler modulus like the sextic $4$-fold, 
\bea
& K =- \log\left[ \widetilde{\cal V}\right]  =- \log\left[ {\cal V} +  i b (t - \bar t) \right], \qquad b= {1 \over 4 \pi^3} \zeta(3) {\alpha'}^3   \int_\M J \wedge c_3, \cr
& K^{t\bar{t}} K_t K_{\bar t}  =  4 \left\{ { 6b - i (t - \bar t)^3  \over 12b + i (t - \bar t)^3  } \right\}^2 = 4 + {144 i b \over  (t - \bar t)^3} + O(b^2), 
\eea
where the volume of $\M$ is given by
\be
{\cal V} = {(t - \bar t)^4\over 4!} \int_\M J \wedge J \wedge J \wedge J. 
\ee
Adding in the effects of fluxes and warping, which are of the same order in $\alpha'$, cannot generically restore the no-scale property since those effects preserve no-scale without accounting for this additional perturbative correction to $K$. 

The second way to interpret this modification of $K$ is via the space-time effective action. Namely, the type IIA string tree-level interactions~\C{typeIIAR4tree}\ generate this modification of the classical $K$. To summarize: for case (A) the no-scale structure is generically broken at tree-level in the string coupling and $O(\alpha'^3)$.   

Case (B) involving M-theory on $\M$ is more problematic. There is no world-sheet analysis available so the only available tool is the space-time effective action. Unlike case (A), there is also no analogue of the string coupling that can suppress the effects of both warping and fluxes. Whether the $R^4$ interactions  of~\C{higherder}\ actually break the no-scale structure is currently unknown despite significant effort~\cite{Grimm:2015mua}. On dimensional reduction, these couplings reduce to the string $1$-loop interactions of~\C{linetwo}. This suggests a possible way of determining whether the no-scale structure is broken by these $R^4$ interactions  
by studying string $1$-loop corrections to the moduli space metric of a $(2,2)$ sigma model with target $\M$. Since the string charge tadpole appears at the same 1-loop order, some form of tadpole cancellation mechanism along with the effects of the warp factor would have to be included for consistency. 

Fortunately the precise order at which the no-scale structure is broken in case (B) is not particularly important for the conclusions we wish to draw. It might play an interesting role in specific moduli stabilization scenarios, but the key observation is that generically the no-scale structure is always broken in even the simplest examples. 

The case of most interest to us is case (C) of F-theory on $\M$. Let us restrict to type IIB orientifold models on a Calabi-Yau 3-fold $\widetilde\M$ for simplicity. Consider type IIB on $\widetilde\M$ with no orientifolding or fluxes initially.  In the notation of~\C{kahler4}, the  K\"ahler potential determined from a $(2,2)$ sigma model with target space $\widetilde\M$ takes the form~\cite{Candelas:1990rm, Honma:2013hma, Halverson:2013qca}:
\be\label{kahler3}
e^{-K(t)} =   -{i\over 3!} \kappa_{ijk}(t^i - \bar t^i) (t^j - \bar t^j) (t^k - \bar t^k)  + {1 \over 4 \pi^3} \zeta(3) {\alpha'}^3  \chi({\widetilde\M}) + O(e^{2\pi i t}).
\ee
The similar form of the $\zeta(3) {\alpha'}^3$ terms in~\C{kahler3}\ and~\C{kahler4}\ is no accident since both the $3$-fold and $4$-fold corrections originate from the same tree-level $R^4$ space-time couplings. For type II string theory, the pure metric $R^4$ couplings of~\C{typeIIBR4tree}\ have been shown to produce precisely the  ${\alpha'}^3$ K\"ahler potential correction of~\C{kahler3}\  for a Calabi-Yau $3$-fold~\cite{Antoniadis:1997eg, Becker:2002nn}. Again this correction breaks the no-scale structure. 

Now we are actually interested in the orientifolded case with fluxes and warping rather than a vanilla compactification on $\widetilde\M$. The extra ingredients make it more likely that the no-scale property is broken at lower order in the $\alpha'$ expansion. There is compelling evidence that this is the case. The first evidence is from a direct computation of string loop corrections to the K\"ahler potential in a toroidal orientifold~\cite{Berg:2005ja}. This accounts directly for the effects of space-time couplings like~\C{Rsquared}. The corrections appear at $O(\alpha'^2)$ suppressed by powers of the string coupling,
\be
K = K_{\rm cl}  + O(\alpha'^2), 
\ee
where $K_{\rm cl} $ is the classical K\"ahler potential accounting for the orientifold projection. 
The second piece of evidence is from the leading quantum corrections to the K\"ahler potential for the heterotic string on a Calabi-Yau $3$-fold, which also appear at $O(\alpha'^2)$ rather than $O(\alpha'^3)$~\cite{Anguelova:2010ed}. The heterotic background is a dual description of type IIB orientifold backgrounds for particular $\widetilde\M$. While moduli-dependence can change when changing duality frames, the dimensionful order of any quantum correction cannot change. 

Despite the breaking of no-scale at $O(\alpha'^2)$, the leading important breaking still happens at $O(\alpha'^3)$. This is because of the observed ``extended no-scale'' structure seen in the D=4 effective field theories obtained from these type IIB backgrounds~\cite{Cicoli:2007xp, Cicoli:2009zh}. The extended no-scale guarantees that any $O(\alpha'^2)$ correction to the K\"ahler potential does not produce a physical potential from~\C{V}\ when $W_0 \neq 0$. Hence we are really interested in the coefficient of the $O(\alpha'^3)$ correction to $K$. 

From~\C{kahler3}, one might conclude that the coefficient of the $O(\alpha'^3)$ correction to $K$ is proportional to the Euler characteristic of $\widetilde\M$, but this is definitely not the case. The O7-planes and D7-branes modify this coefficient in a way that has been examined in recent work~\cite{Minasian:2015bxa}. This analysis is not the full story because the answer is not a quantity expressed in terms of the geometry of the underlying $4$-fold $\M$, and warping has not been taken into account, but it does show that the coefficient is modified.  The important observation for us is that for case (C), the breaking of the no-scale structure which generates a physical potential is at tree-level in the string loop expansion and at $O(\alpha'^3)$ in the derivative expansion.

\subsection{The punch line}

We can now get to the main observation quite quickly. The effective field theory data given in~\C{basicdata}\ is misleading for a generic target space, even with the instanton correction omitted. The superpotential obtained from expanding around a supersymmetry breaking flux configuration is indeed $W=W_0$ and protected from perturbative renormalization, but the K\"ahler potential can never be truncated to its no scale form in any approximation for a generic target space. By doing so, it would appear that the supersymmetry breaking background solves the space-time equations of motion when, in reality, there is a physical potential proportional to $|W_0|^2$ and no static solution. This physical potential should be viewed as a {\it classical} potential reflecting the failure to solve the space-time equations of motion. It is a potential that generically involves all the moduli. 

The corrections to the K\"ahler potential from higher derivative terms are in no sense unimportant because there is no compact supergravity  flux solution about which to expand. This is critical because quantum corrections must be computed around a given background. The classical picture for the case with $W_0\neq 0$ looks schematically more like figure~\ref{figure2}\ rather than figure~\ref{figure1}. In understanding the structure of quantum corrections for a background with $W_0\neq 0$, there are a number of issues that need to be addressed. 

\subsubsection*{\ul{\it String theory versus effective field theory}}

It is helpful to contrast three type IIB backgrounds: the supersymmetry preserving flux model with $W_0 = 0$, the supersymmetry breaking model with $W_0 \neq 0$, and compactification on an $n$-dimensional sphere of radius $R$. Neither of the last two models is a solution to the space-time equations of motion. In both latter cases, the potential is small at large volume. For the sphere, the effective potential generated from curvature in Einstein frame takes the form,
\be
V_{eff}  =  - {n (n-1) \over R^{16\over 8-n}}, 
\ee
which drives the sphere to small volume. The basic intuition for both backgrounds that generate a space-time potential is that the classical potential dominates in the weak coupling region. This is essentially what it means to have a semi-classical expansion. For example, the sphere should time evolve to smaller radius and larger curvature until a semi-classical analysis is no longer trustworthy. 

The flux background with $W_0 \neq 0$ has similarities and differences with the sphere compactification. Both have a potential and that potential can be made arbitrarily small by making the volume arbitrarily big. The potential for the flux background is generated by higher derivative terms and so is more suppressed by powers of $\ell_p$, while the sphere potential comes directly from curvature at the two derivative level. Although $W_0$ is not tunable and often taken to be moderately large in many scenarios, like the LARGE scenario where $W_0$ is $O(1-100)$, there are examples where small values can be found; see, for example~\cite{MartinezPedrera:2012rs}. Small choices for $W_0$ are typically assumed in the KKLT scenario. The real issue is whether the smallness of the potential at some fixed volume for $\widetilde\M$ means anything since there is no static background, or whether it simply sets a time scale for the applicability of a weak coupling analysis. 

To answer this question, we require a framework for computing quantum corrections. String theory should be that framework but we immediately face an obstacle. String theory requires an on-shell solution. There is no currently understood method of computing quantum corrections off-shell, and this is closely tied to the fact that observables in string theory are always correlation functions of vertex operators, which are only defined in an on-shell background. So the first question is whether flux backgrounds with $W_0 \neq 0$ can even be promoted to time-dependent solutions of the space-time equations of motion. 

From an N=1 D=4 perspective, this might appear possible by assigning appropriate time evolution to the K\"ahler moduli. However, that is simply masking the difficulty of finding a D=10 time-dependent solution of string theory. The basic issue is whether a choice of metric for $\widetilde\M$, fluxes and suitable time-dependence can satisfy the gauge constraints of the type IIB space-time effective action; namely the higher-dimensional and higher derivative analogues of the familiar Hamiltonian and momentum constraints of general relativity. These constraints are equivalent to solving the following subset of the space-time equations of motion,
\be
G_{0\mu} = 8\pi T_{0\mu}, 
\ee
where $\mu$ runs over the non-compact space-time directions.  
If so then this data can specify a Cauchy surface for some cosmological space-time. This is already a difficult question to answer, but one thing seems clear: either in the far past or far future, the background will be strongly coupled.  

Without an on-shell solution, there is no recipe for even computing string loop corrections to the effective potential. However, we might imagine ignoring string theory and simply studying these flux configurations as off-shell backgrounds of the space-time effective action. This does not alleviate the difficulty in solving the gauge constraints that must be satisfied by good initial value data in any theory of gravity, but it does provide a potential framework for computing loop effects, using some choice of Planck scale cutoff and a cutoff in time. 

We might imagine a slowly rolling solution at large volume with sufficiently slow time-dependence that four-dimensional loop effects might reliably be computed. For example for computing corrections to the space-time potential from loops of massive gravitinos. This appears promising as a means of stabilizing the K\"ahler moduli; for work on this direction, see~\cite{Berg:2005yu}. Intuitively, such an approach should be sensible in effective field theory if the time-dependence is sufficiently slow that it can be ignored; if so, then one might imagine following a similar strategy in Lorentzian time-dependent string theory. Understanding how to compute such loop effects in string theory around an off-shell configuration is an interesting issue. 

There is a nice tractable field theory example of this type which consists of two-dimensional supersymmetric Yang-Mills on a Milne space~\cite{Craps:2006xq}. Supersymmetry is only broken by the time-dependence of the Milne metric which takes the form,
\be
ds^2 = e^{2Qt} \left( -dt^2 +d\s^2 \right), 
\ee 
with $Q$ a fixed parameter and $\sigma$ a periodic spatial coordinate. In the far future $t\rightarrow \infty$, supersymmetry is restored. The Coulomb branch effective theory at $1$-loop is computed by integrating out massive W-bosons with time-dependent masses that depend on a Coulomb branch modulus $b$ as follows: 
\be
m_W^2 \sim e^{2Q t} b^2. 
\ee
The result is an explicit time-dependent potential that decays exponentially rapidly in $ e^{Qt}$  at late times. This potential is only generated because of the metric time-dependence. In a static supersymmetric background, the potential vanishes. While this model has a fixed background metric and no dynamical gravity, the form of the generated potential at least supports the notion that perturbative potentials can be computed for very slowly rolling backgrounds by ignoring the time-dependence, up to very small time-dependent corrections. 

\subsubsection*{\ul{\it Instanton effects}}

The biggest issue, however, is the question of instanton or non-perturbative effects. Specifically the effects that can generate the non-perturbative terms in $W$ of~\C{basicdata}.  In static backgrounds, these effects are not generated by generic Euclidean instantons, but by BPS configurations. These quantum corrections are the essential ingredient for the majority of string landscape models. 

Let us start with the better understood case of a supersymmetric flux vacuum with $W_0=0$. Already in this case, there is much yet to be understood about instanton corrections. The most basic issue is that the instantons of flux backgrounds are much closer cousins of non-perturbative effects in Chern-Simons theories than Yang-Mills theories. 

To see this, consider models in class (B) consisting of M-theory on $\M$ discussed in section~\ref{higherderint}. The action has two Chern-Simons couplings
\be\label{complexify}
 - {1\over 12\kappa_{11}^2} \int C_3\wedge G_4 \wedge G_4 - T_{M2} \int C_3 \wedge X_8.
\ee
Tadpole cancelation guarantees that the variation with respect to $C_3$ vanishes on integration over $\M$. However, a non-zero $G_4$ generically leads to non-zero Chern-Simons couplings for the three-dimensional vector multiplets that arise from Kaluza-Klein reduction on $\M$. From the three-dimensional perspective, Euclidean instantons correspond to abelian instantons for these gauge-fields.  
This means that the Euclidean action, whose saddle points in BPS situations usually correspond to instantons, is generically complex not real. This suggests that instantons in flux vacua do not look like simple Euclidean M5-branes wrapping appropriate divisors of $\M$, but that complex field configurations are needed to describe non-perturbative saddle points.\footnote{It is worth emphasizing that it is not simply that~\C{complexify}\ complexifies the action. A non-vanishing $\theta$-angle in D=4 Yang-Mills also complexifies the action, but does not require complex field configurations. Rather~\C{complexify}\ contributes to the equations of motion for $C_3$, which is why complex field configurations are needed. I would like to thank Ilarion Melnikov for suggesting this clarification. } This is a fascinating issue but one that requires a separate analysis.  Note that with $W_0=0$, there is no reason to expect any interesting new vacua generated by instanton corrections at large volume. The generic expectation is runaway behavior. This same issue of complex field configurations also arises in models of class (A) and class (C). In each case, there are activated Chern-Simons interactions that complexify the Euclidean action. 

Now we should ask about models with $W_0 \neq 0$. For such models, there are most certainly non-perturbative corrections to the space-time {\it potential}, but they are hard to understand in any framework. In string theory, wrapped branes require an on-shell background for kappa symmetry. So one is forced to analyze non-perturbative saddle points of a time-dependent background. This kind of question has not been addressed even in field theory, let alone quantum gravity. It is not even meaningful to ask about non-perturbative corrections to the space-time potential without specifying data like initial conditions, which define the classical time-dependent background. 

The situation, however, is more problematic. 
In the time-dependent setting with no preserved supersymmetry, it is not even clear how to formulate an analogue of the usual Euclidean BPS instanton condition that 
\be
\delta \Phi =0,  
\ee 
where $\Phi$ denotes the fields of the theory and $\delta$ denotes an unbroken supersymmetry generator.  
It can no longer be a statement uniform in the volume parameter of the compactification space $\M$ or $\widetilde\M$, depending on the model under consideration. When the volume is large, there is low scale supersymmetry breaking and we might expect some notion of BPS to exist, at least in the static case. However, when the volume is small supersymmetry is broken at a high scale relative to the Kaluza-Klein scale and there is no notion of low-energy supersymmetry. 

A reasonable working intuition is that any non-perturbative corrections to the space-time potential should not dominate 
the physics at large volume, which is classically rolling. That would almost amount to a violation of the premise of a semi-classical expansion. If the classical potential is small then loop corrections might play an important role. This should be contrasted with instanton effects in a static background for which a non-perturbative potential might well generate the leading effect. We will see some evidence that this intuition is correct momentarily.


The other possible source of a non-perturbative contribution to $W$ is from strong coupling gauge dynamics like gaugino condensation on some stack of branes; typically either D7-branes or D3-branes. This case may look superficially different from the computation of Euclidean instantons but it should be closely related. The robust way to compute the contribution to the superpotential from strong gauge dynamics is to compactify space-time  
\be
\R^4 \, \rightarrow \, \R^{3}\times S^1, 
\ee
and turn on a Wilson line on the $S^1$ to go to the Coulomb branch of the $D=3$ gauge theory. The superpotential is then generated by the contribution of abelian Euclidean instantons. This approach, which was studied from brane dynamics in~\cite{Lee:1997vp}, has been successfully used to compute F-theory superpotentials in~\cite{Katz:1996th}. The point for us is that all the issues that arise for abelian instantons should also be present for strong gauge dynamics. 

Another way to think about problems with strong gauge dynamics in a time-dependent background is to consider a slowly rolling large volume solution and imagine that gaugino condensation occurs at some late time generating a superpotential. The first issue would be at what time? The gauge coupling for wrapped branes will typically become weaker at late times as volumes expand so the immediate issue would be why the condensation had not already occurred in the far past, where volumes were small. The other puzzle would be why the asymptotic vacuum energy of the universe would change as opposed to a local dynamical response to the strong dynamics of the gauge theory since the gauge degrees of freedom are always supported on localized branes.

These puzzles all appear to be reflections of the problems that arise in trying to generate a superpotential contribution in a non-static background. It would be very nice to develop a theory of non-perturbative corrections in time-dependent or off-shell backgrounds. This would help provide a more robust understanding of whether strong gauge dynamics or other instanton effects might actually have an effect on the vacuum structure of these backgrounds analogous to BPS instantons in the static case. 

\subsubsection*{\ul{\it Implications of the heterotic no-go result}}

Indeed, the only precise data we currently possess about supersymmetry breaking and instanton effects comes from the heterotic string. The evidence suggests that 
 the interplay between $W_0$ and instantons does not generate new metastable vacua; in this case specifically de Sitter vacua. 
 
 Many type IIB flux orientifold models can be dualized to heterotic backgrounds~\cite{Dasgupta:1999ss, Becker:2009df, Becker:2009zx}.  If there is a flux-generated $W_0$ of any size in the type IIB model then a combination of NS flux and geometry should generate the same $W_0$ in the heterotic dual. Indeed there is a dual superpotential proposed for heterotic flux backgrounds taking the form~\cite{Becker:2003gq, LopesCardoso:2003dvb}
\be
W  = \int  \left( {\cal H} + i dJ \right)\wedge  \Omega_3,
\ee
where $\Omega_3$ is the holomorphic $3$-form,  $ {\cal H}$ is the gauge-invariant heterotic flux and $J$ is the fundamental form for the compactification space; see~\cite{Becker:2003dz} for an overview. 
It is very useful to think about the physics in the heterotic frame because all the effects of fluxes are visible at tree-level in the string coupling so conformal field theory becomes a powerful tool. 

There are only three classes of non-perturbative corrections that can renormalize a superpotential: those generated by tree-level world-sheet instantons, those generated by space-time NS5-brane instantons and those generated by strong gauge dynamics. The former are completely captured by the tree-level world-sheet conformal field theory. From an effective field theory perspective, we should expect a generic tree-level  superpotential just like~\C{basicdata}\ of the form
\be\label{hetsuper}
W=W_0 + A e^{-a t},
\ee
where $t$ are K\"ahler moduli. The non-perturbative terms in~\C{hetsuper}\ are generated by world-sheet instantons.   If a scenario like the K\"ahler uplift scenario of section~\ref{intro}, which is based on the structure of generic quantum corrections, were possible then we might expect some de Sitter critical points of this potential for some model. This should be true at tree-level in the heterotic string, regardless of whether string loop effects or string non-perturbative effects generate additional potentials, or even destabilize the dilaton. It is just a consequence of the genericity arguments used in landscape constructions. Yet no-go theorem (iii) rules out this possibility, and even rules out de Sitter critical points for backgrounds with large curvature! 

The reason should now be clear. The generic weak coupling behavior of models with $W_0 \neq 0$ is not static but rolling, assuming such models can even be promoted to consistent time-dependent backgrounds. There is no reason to expect a stabilized macroscopic dS solution. 
This picture resolves a number of puzzles. It  resolves all the tension with the no-go theorems of section~\ref{intro}. It also resolves puzzles with dualizing non-supersymmetric type IIB backgrounds to heterotic backgrounds, which subsequently fail to satisfy the space-time equations of motion~\cite{Becker:2009zx}. There might still be interesting fully stabilized vacua with $W_0 \neq 0$, but they are more likely to involve loop contributions to the space-time potential rather than non-perturbative effects. The computation of such loop effects around an off-shell, or equivalently an on-shell time-dependent, configuration in both string theory and effective field theory needs to be understood more deeply.

\section*{Acknowledgements}

I would like to thank the following people for helpful discussions and comments without implying any concordance with my conclusions: Katrin Becker, Melanie Becker,  Michele Cicoli, Joseph Conlon, Keshav Dasgupta, Sergei Gukov, Luca Martucci, Liam McAllister, Evan McDonough, Juan Maldacena, Travis Maxfield, Ilarion Melnikov, Ruben Minasian, Fernando Quevedo, Callum Quigley, Thomas Van Riet, Daniel Robbins, Bob Wald and Alexander Westphal. 

I would also like to thank the organizers of the Theoretical Cosmology Meeting at Leuven and the GRC meeting on String Theory and Cosmology in Tuscany. Lastly, I would particularly like to thank Joe Polchinski for a past lively discussion which led to this work. 
S.~S. is supported, in part, by NSF Grant No.~PHY-1720480. 





\newpage

\providecommand{\href}[2]{#2}\begingroup\raggedright\endgroup


\end{document}